\documentclass[12pt,preprint, nofootinbib, aps]{revtex4}
\usepackage{graphicx}
\usepackage{dcolumn}
\usepackage{bm}
\usepackage{epsfig}
\usepackage{amssymb,amsmath,amsfonts,amsbsy,epsfig}
\usepackage{color}

\def\lsim{\mathrel{\rlap{\lower4pt\hbox{\hskip1pt$\sim$}}
    \raise1pt\hbox{$<$}}}         
\def\gsim{\mathrel{\rlap{\lower4pt\hbox{\hskip1pt$\sim$}}
    \raise1pt\hbox{$>$}}}
\def\fun#1#2{\lower3.6pt\vbox{\baselineskip0pt\lineskip.9pt
  \ialign{$\mathsurround=0pt#1\hfil##\hfil$\crcr#2\crcr\sim\crcr}}}

\input epsf

\def\beq{\begin{equation}}
\def\eeq{\end{equation}}
\def\bea{\begin{eqnarray}}
\def\eea{\end{eqnarray}}

\def\nn{\nonumber}

\newcommand{\degree}{\ensuremath{^\circ}}

\begin{document}

\preprint{IPMU-09-0135}
\title{Axigluon as Possible Explanation for $p\bar{p} \rightarrow t\bar{t}$ Forward-Backward Asymmetry}
\author{Paul H. Frampton~$^{\bf a,b}$}
\email{paul.frampton@ipmu.jp}
\author{Jing Shu~$^{\bf b}$}
\email{jing.shu@ipmu.jp}
\author{Kai Wang~$^{\bf b}$}
\email{kai.wang@ipmu.jp}
\affiliation{
$^{\bf a}$ Department of Physics and Astronomy, University of North Carolina,
Chapel Hill, NC 27599-3255, USA.\\
$^{\bf b}$Institute for the Physics and Mathematics of the Universe, University of Tokyo, Kashiwa, Chiba 277-8568, JAPAN
}

\begin{abstract}
A flavor-nonuniversal chiral color model is introduced.
It is used for comparison to the recent
data on $\bar{p} p \rightarrow \bar{t} t$. 
We concluded that the data are consistent with
interpretation as an axigluon exchange within 1$\sigma$ and a unique rise and fall behavior 
is predicated with regard to the asymmetry $A^t_{FB}$ as a function
of $t \bar{t}$ invariant mass, which can distinguish our model from others before one discovers the axigluon resonance.  
Further aspects of the model are discussed.

\end{abstract}
\maketitle

\section{Introduction}

Since it underlies nuclear physics and hadronic physics, it is important to understand whether the standard version of the theory of strong interactions, quantum chromodynamics (QCD), is an adequate description at much higher energies. For example, although electroweak interactions exhibit parity violation, QCD theory as presently formulated respects parity conservation.

One generalization of QCD proposed some time ago is 
chiral color \cite{PS, FG} in which the color gauge group arises 
from the spontaneous breaking of a larger group at higher energy. This gives rise to a massive
color octet of gauge bosons, axigluons, which acquire mass through the Higgs mechanism.
The earliest chiral color models treated the different quark flavors
universally. A flavor-nonuniversal version which, however, involved extension of the
electroweak gauge group was discussed in \cite{PF}.
In the present article we introduce a new flavor-nonuniversal chiral
color model in which the electroweak symmetry is unchanged from the standard model,
thus is more conservative, and seek within it a unique experimental signature
below the energy of the axigluon resonance.

One possible relevant experiment concerns the study of heavy quark asymmetries. Both CDF and D0 at Tevatron have observed large forward-backward asymmetry values in top anti-top pair events ($A^t_{FB}$)  \cite{cdf,newcdf,d0}. The latest measurement of $A_{FB}^t$ is from the CDF experiment where they obtained:
\beq
A_{FB}^t = 0.193 \pm 0.065_{\, \rm stat.} 
\pm 0.024_{\, \rm syst.}~
\label{eq:newcdf}
\eeq 
in the lab frame with $3.2$~fb$^{-1}$ integrated luminosity data~\cite{newcdf}.
Within the standard model (SM),
the next-to-leading-order (NLO) QCD interference effects predict $A_{FB}^t(SM) = 0.051$ ~\cite{Antunano:2007da,Bowen:2005ap,mynlo} and
it is stable with respect to the QCD threshold resummation \cite{Almeida:2008ug} 
\footnote[1]{In the
$t\bar{t}+$jet production, however, a large cancellation of the top forward backward asymmetry occurs from NLO corrections \cite{Dittmaier:2007wz}.}. This is not enough to account for the measured large asymmetry and suggests there might be new physics in the top quark sector.  

There have been several new physics proposals trying to 
explain this large positive asymmetry. However, any model trying to 
explain the large asymmetry encounters a challenge as it must also
avoid predicting large deviation in total cross section 
$\sigma_{t\bar{t}}$ and $t\bar{t}$ invariant mass $M_{t\bar{t}}$ 
since they are in good agreement with the SM predictions 
\footnote[2]{We notice that the $t\bar{t}$ invariant mass $M_{t\bar{t}}$ 
distribution measured at the Tevatron is slightly softer than
SM prediction.}. 
One category of model suggests new physics in the $t$-channel 
with large flavor violating $ut$ or $dt$ couplings. Phenomenologically 
viable examples include $t$-channel $Z'$/$W'$ exchange with 
large flavor violating couplings \cite{Jung:2009jz,Cheung:2009ch}, 
$t$-channel flavor and fermion number violating scalar as $SU(3)_{C}$ 
triplet or sextet\cite{tim}. Nevertheless, it is difficult to 
imagine how such {\it ad hoc} couplings can be naturally generated.

Another category of model involves vector bosons in the 
$s$-channel with chiral coupling to the light quarks and the top quark.
A color octet particle is preferred in order to use its large QCD interference 
effect.  $\sigma_{t\bar{t}}$ and $M_{t\bar{t}}$ also constrain 
the mass of the resonance.  
The top forward-backward 
asymmetry requires that parity transformation is 
violated if we apply it to the final top quark pair. 
Similarly, parity transformation applying to the initial 
light quark pair must also be violated which can be easily seen 
with an additional 180 degree rotation along the scattering 
plane.
Since the QCD interactions already provide the vector couplings 
for the interference, a spin one, color octet with nonzero axial couplings 
is the basic requirement.  

Such a new particle can be realized within 
a wide class of theories,  invoked for 
phenomenological reasons \cite{FG} inspired by the
weak interactions, used to explain the electroweak 
symmetry breaking \cite{Hill:1991at}, or motivated 
by study of extra dimensions \cite{Lillie:2007yh, Agashe:2006hk}. 
It has previously been observed by Ref. \cite{Ferrario:2009bz}
that only the $g^{q}_{A} g^{t}_{A} < 0$ case gives rise to 
a positive asymmetry from interference and we pursue that here. In this paper, we 
consider a simple extension of the chiral model involving a fourth
family and based on only two extra parameters to explain 
the observations. It leads to a striking prediction for a \textit{rise and fall} 
behavior of $A^t_{FB}$ as a function of $M_{t \bar{t} }$
as one approaches the axigluon resonance.

\section{Chiral Color with a Fourth Family}

We consider a chiral color model based on gauge group $SU(3)_A \times SU(3)_B 
\times SU(2)_L \times U(1)_Y$ with gauge couplings 
$g_A$, $g_B$, $g_L$ and $g_Y$ respectively. In this model, quarks of opposite chiralities are charged 
under different $SU(3)$ gauge groups. Similarly, quarks with the same chiralities but between the
first two generations and the third and fourth generations
(for reviews, see {\it e.g.}\cite{FHS}) \footnote[3]{At the low energy, the fourth generation appears as a Wess-Zumino-Witten term, which introduces one-loop suppressed counter terms among the massive octet gauge boson and QCD gluon.}
are charged under different $SU(3)$ gauge groups. 
There is a bi-triplet scalar field $\Sigma$ which gets a vev at the TeV scale 
$ \langle \Sigma_{i\bar{k}} \rangle = u \delta_{i\bar{k}}$ and 
breaks the $SU(3)_A$ and $SU(3)_B$ to the diagonal 
subgroup, the QCD color group $SU(3)_c$. 
The SM Higgs scalars are now split into two parts, the quarkonic Higgs $H_q$ which is a triplet under color groups $SU(3)_A$ and $SU(3)_B$ and the leptonic Higgs $H_l$ which is a singlet under color groups. The full field contents are summarized in Table \ref{table:assign}. 

\begin{table}[h]
\begin{center}
{\renewcommand{\arraystretch}{1.1}
\begin{tabular}{ c || c  c  c|  c  c  c |c  c|c c c }
  \hline
  \rule[5mm]{0mm}{0pt}
Field & $Q_i$ & $u^{c}_i$ & $d^{c}_i$ & $Q_j$ & 
$u^{c}_j$ & $d^{c}_j$ &  $\Sigma$ & $H_{q}$ & $L_k$ & $e^c_k$ & $H_{l}$ \\
\hline
\rule[5mm]{0mm}{0pt}
SU(3)$_A$ &  \bf{3} & \bf{1} & \bf{1} & \bf{1} & 
$\bar{\bf{3}}$ &  $\bar{\bf{3}}$ &  \bf{3} &  \bf{3} & \bf{1} & \bf{1} & \bf{1}\\
 \rule[5mm]{0mm}{0pt}
SU(3)$_B$ &  \bf{1} & $\bf{\bar{3}}$ & $\bf{\bar{3}}$ & 
\bf{3} &  \bf{1} &  \bf{1} &  $\bf{\bar{3}}$  & 
$\bf{\bar{3}}$ & \bf{1} & \bf{1} & \bf{1} \\
 \rule[5mm]{0mm}{0pt}
SU(2)$_{L}$ &  \bf{2} &  \bf{1} &  \bf{1} & \bf{2} &  
\bf{1} &  \bf{1} &  \bf{1} &  \bf{2} & \bf{2} & \bf{1} & \bf{2} \\
 \rule[5mm]{0mm}{0pt}
U(1)$_{Y}$ & ${1/ 3}$ & $-{4/ 3}$ & ${2/3}$ &  
${1/3}$ & $-{4/3}$ & ${2/3}$ & 0 & 1 & -1 & 2 & 1\\
\hline
\end{tabular}
}
\caption{Charge assignment of all the quark, lepton fields 
and the Higgs fields under $SU(3)_A\times SU(3)_B\times 
SU(2)_L\times U(1)_Y$. The flavor indices are as 
$i=1,2$, $j=3,4$ and $k=1,2,3,4$. The hypercharge is 
defined as the convention of electric charge 
$q= I^3_L+Y/2$ where $I^3_L$ is the third component 
of $SU(2)_L$ isospin. All the $SU(2)_L$ singlet fields 
are defined in their conjugate forms.}
\label{table:assign}
\end{center}
\end{table}

The kinetic term for the link field becomes the 
mass term for the massive gauge boson 
$
\rm{Tr}[(D_{\mu} \Sigma)^{\dag} (D_{\mu} \Sigma)] 
\supset {u^2} ( g_A A_{\mu} - g_B B_{\mu})^2 /2
 = {u^2 g^2} (G_{\mu}^{1})^2 /2 
$, where $g \equiv \sqrt{g_A^2 + g_B^2} $. 
The rotation matrix between gauge bosons in mass eigenstates
and gauge eigenstates is
\begin{eqnarray}
\begin{pmatrix}
G^1_{\mu} \\ G^0_{\mu}
\end{pmatrix}
=
\begin{pmatrix}
s_g&-c_g\\
c_g&s_g
\end{pmatrix} 
\begin{pmatrix}
A_{\mu} \\ B_{\mu}
\end{pmatrix}
\ ,
 \label{G_MG}
\end{eqnarray}
where we define $s_g \equiv \sin\theta \equiv  g_A/g$ and 
$c_g \equiv  \cos \theta = g_B/g$ so 
$\theta = \arctan (g_A/g_B)$. The massless 
field $G^0_\mu$ is the usual QCD gluon while we call 
the massive octet vector boson $G^1_\mu$ ``axigluon".  

The QCD couplings are flavor universal as $g_s = g s_g c_g$. There is a redundant symmetry of the model $g_A \leftrightarrow g_B$, so only half of the parameter space in $\theta$ is physical ($\theta \leftrightarrow 90\degree - \theta$). Requiring that our model is perturbative $g_A, g_B < 2 \pi$ we impose
a further constraint $10\degree < \theta < 45 \degree$.
For the fermions charged under gauge group $SU(3)_A$ 
and $SU(3)_B$, their couplings to the massive 
axigluon $G^1_\mu$ are $gs_g^2$ and $-g c_g^2$ 
respectively. Therefore, the vector and axi-vector coupling strengths are   
\beq
g^{q}_{V} = g^{t}_{V} = - \frac{g{c_{2g}}}{2},~~- g^{q}_{A} =  g^{t}_{A} = \frac{g}{2}
\eeq 
where $c_{2g} \equiv \cos (2 \theta)$. We can see that the axial-vector couplings are always nonzero and the above coupling strengths satisfy the following relations 
\bea
\label{eq: condition}
g_A^q g_A^t <0, ~~g_V^q g_V^t >0~~~  \mathrm{with }~~~ g_V < g_A \ ,
\eea which are crucial for the phenomenology predictions below. 

\section{Forward-Backward Asymmetry}

The $V-A$ structure in axigluon interaction will induce
parity violation in $q\bar{q}\to t\bar{t}$ and consequently a 
non-vanishing forward-backward asymmetry occurs in such process at the $p\bar{p}$ 
collider. To see this feature analytically, in Eq. (\ref{eq:qqtt}), we give the matrix element square for $q\bar{q}\to t\bar{t}$  
of the two leading production channels as the $s$-channel gluon and axigluon. $\gamma/Z$ contribution
do not interfere with the gluon diagram and are very small thus can be neglected.
\bea
&& \sum \left |{\cal M} \right |^2  =
g^4_{s}(1+c^2+4m^2)   \nn \\ && + \frac{2 g^{2}_{s}\hat{s} (\hat{s}-M_G^2)}
{(\hat{s}-M_G^2)^2+M_G^2 \Gamma_G^2}
\left[ g_V^q \, g_V^t \, (1+c^2+4m^2) \right. \nn \\ && \left. 
+ 2 \, g_A^q \, g_A^t \, c  \right] +
\frac{\hat{s}^2} {(\hat{s}-M_G^2)^2+M_G^2 \Gamma_G^2}
\left[ \left( (g_V^q)^2+(g_A^q)^2 \right) \right. \nn \\ && \times 
\left( (g_V^t)^2 (1+c^2+4m^2) +  (g_A^t)^2 (1+c^2-4m^2) \right)
\nn \\ && \left.
+ 8 \, g_V^q \, g_A^q \, g_V^t \, g_A^t \, c \, \right]
~,
\label{eq:qqtt}
\eea
where $m=m_t/\sqrt{\hat{s}}$, $\beta = \sqrt{1-4m^2}$ is the velocity of the top quark
 and $c = \beta \cos\theta$ where $\theta$ is the polar angle of the 
top quark with respect to the incoming up quark in the parton center-of-mass frame.

The asymmetry $A^{t}_{FB}$ can be defined using the total number of
top quark in positive and negative semi-sphere as
\beq
A^{t}_{FB}= \frac{N_{t}(\eta\ge 0)-N_{t}(\eta\le 0)}{N_{t}(\eta\ge 0)+N_{t}(\eta\le 0)},
\eeq
where is $\eta$ is the rapidity of the top quark. Under CP transformation, there also exists a
equivalent definition by changing $N_{t}(\eta\le 0)$ to $N_{\bar{t}}(\eta\ge 0)$. 
We can qualitatively see the asymmetry behavior from the linear term $(1\pm \cos\theta)$
in the matrix element squared where $(1+\cos\theta)$ corresponds to the maximal positive asymmetry.
Equation (\ref{eq:qqtt}) shows the correlation between $ \cos\theta$ and the relative signs 
of the couplings. The contribution to $A^{t}_{FB}$ due to the interference term
depends on the sign of $g^{q}_{A}g^{q}_{A}$.
The new physics contribution depends on the sign of 
$g^{q}_{V}g^{t}_{V}g^{q}_{A}g^{t}_{A}$. 
The key prediction of our model is that $g^{q}_{V}=g^{t}_{V}$ 
while $g^{q}_{A}=-g^{t}_{A}$. Therefore, both $g^{q}_{A}g^{q}_{A}$ and $g^{q}_{V}g^{t}_{V}g^{q}_{A}g^{t}_{A}$ are negative and the interference term induces a positive asymmetry while the new physics term induces a negative asymmetry below the resonance. Since most the events are far below resonance, the interference term completely overwhelms the new physics term when counting the overall $A_{FB}^t$. Nevertheless, at higher $M_{t \bar{t}}$ 
invariant mass, we may expect a competition between the interference term
and the new physics term to provide an interesting 
signature to distinguish our model.

\section{Numerical Results}

The SM prediction of $\sigma_{t\bar{t}}$ is in 
good agreement with measurements from CDF 4.6~fb$^{-1}$ luminosity data \cite{cdf-ttbar} 
\beq
\sigma^{CDF} (t \bar{t}) =  7.50 \pm 0.31_{\mathrm{stat}} \pm 0.34_{\mathrm{syst}}  \pm 0.15 _{\mathrm{Z ~ theory} } ~ \mathrm{pb}  \ ,
\eeq
where the top quark mass is assumed to be 172.5 GeV. This CDF result is consistent with measurements from
D0 \cite{Abazov:2009ae} that use smaller data sets. 
We combine different errors in quadrature and obtain $\sigma^{exp} = 7.50 \pm 0.48 $ pb.

The $t \bar{t}$ invariant mass distribution was also measured recently by CDF  
\cite{Aaltonen:2009iz}, the data from the last bin which are expected to give the most stringent constrain on our model are 
\begin{eqnarray}
&& \frac{d\sigma}{dM_{t\bar t}} (0.8-1.4~{\rm TeV})  = \nonumber \\
&&0.068 \pm 0.032_{\, \rm stat.} \pm 0.015_{\, \rm syst.}   
\pm 0.004_{\, \rm lumi.}~\text{fb/GeV}~.
\label{mttbar}
\end{eqnarray}
We combine different errors in quadrature as $d\sigma/dM_{t\bar
 t}  (0.8-1.4~{\rm TeV}) = 0.068 \pm 0.036$ (fb$/$GeV).
 
\begin{figure}[t]
\includegraphics[width=14cm]{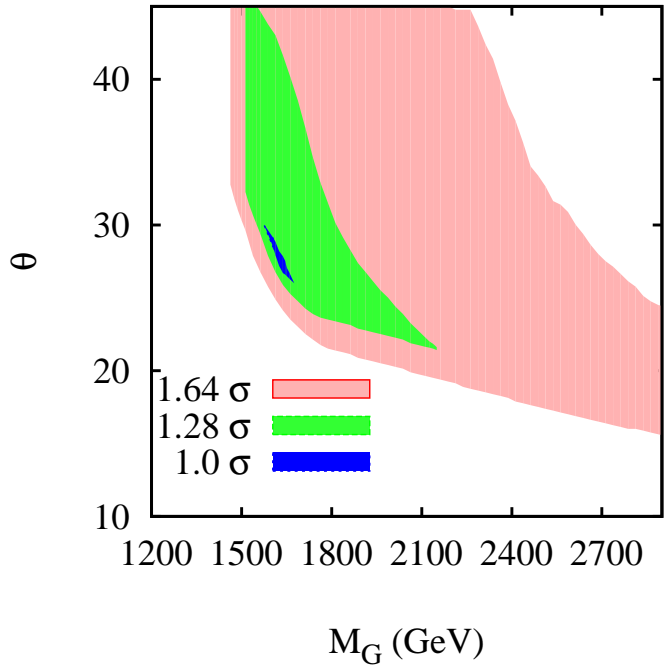}
\caption{Contour plot for $1~\sigma$, $1.28~\sigma$ and $1.64~\sigma$
allowed parameter space in the $\theta$ - $M_G$ plane.}
\label{Fig:contour}
\end{figure}

 Our simulation is only a leading order calculation. 
 To fit the total cross section, we have multiplied by
 a NLO $K$-factor as $K=1.329$ for $m_{t}= 172.5$~GeV.
 To be consistent with measurements, 
 the $A^{t}_{FB}$ and $M_{t\bar{t}}$ results are based on
 simulation using $m_{t}=175$~GeV. However since QCD corrections
 often soften the $M_{t\bar{t}}$ at high value, 
 we did not multiply any $K$ factor for our simulated
 result of the last bin value in $M_{t\bar{t}}$.
 Taking into account all the above constraints, we scan the allowed parameter space in terms
of the axigluon mass $M_G$  and the mixing angle $\theta$ between gauge groups $SU(3)_A$ and $SU(3)_B$ in Fig. \ref{Fig:contour}. The $1~\sigma$, $1.28~\sigma$ ($80\%$ C.L.)  and $1.64~\sigma$ ($90\%$ C.L.) regions are
labelled accordingly. 

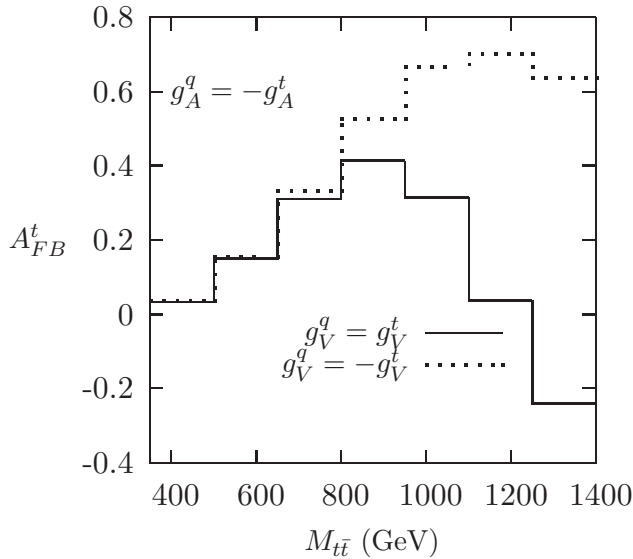
\begin{figure}[t]
\setlength{\unitlength}{0.240900pt}
\ifx\plotpoint\undefined\newsavebox{\plotpoint}\fi
\sbox{\plotpoint}{\rule[-0.200pt]{0.400pt}{0.400pt}}%
\begin{picture}(1500,900)(0,0)
\font\gnuplot=cmr10 at 12pt
\gnuplot
\sbox{\plotpoint}{\rule[-0.200pt]{0.400pt}{0.400pt}}%
\put(225.0,150.0){\rule[-0.200pt]{4.818pt}{0.400pt}}
\put(200,150){\makebox(0,0)[r]{-0.4}}
\put(905.0,150.0){\rule[-0.200pt]{4.818pt}{0.400pt}}
\put(225.0,267.0){\rule[-0.200pt]{4.818pt}{0.400pt}}
\put(200,267){\makebox(0,0)[r]{-0.2}}
\put(905.0,267.0){\rule[-0.200pt]{4.818pt}{0.400pt}}
\put(225.0,383.0){\rule[-0.200pt]{4.818pt}{0.400pt}}
\put(200,383){\makebox(0,0)[r]{ 0}}
\put(905.0,383.0){\rule[-0.200pt]{4.818pt}{0.400pt}}
\put(225.0,500.0){\rule[-0.200pt]{4.818pt}{0.400pt}}
\put(200,500){\makebox(0,0)[r]{ 0.2}}
\put(905.0,500.0){\rule[-0.200pt]{4.818pt}{0.400pt}}
\put(225.0,617.0){\rule[-0.200pt]{4.818pt}{0.400pt}}
\put(200,617){\makebox(0,0)[r]{ 0.4}}
\put(905.0,617.0){\rule[-0.200pt]{4.818pt}{0.400pt}}
\put(225.0,733.0){\rule[-0.200pt]{4.818pt}{0.400pt}}
\put(200,733){\makebox(0,0)[r]{ 0.6}}
\put(905.0,733.0){\rule[-0.200pt]{4.818pt}{0.400pt}}
\put(225.0,850.0){\rule[-0.200pt]{4.818pt}{0.400pt}}
\put(200,850){\makebox(0,0)[r]{ 0.8}}
\put(905.0,850.0){\rule[-0.200pt]{4.818pt}{0.400pt}}
\put(258.0,150.0){\rule[-0.200pt]{0.400pt}{4.818pt}}
\put(258,100){\makebox(0,0){ 400}}
\put(258.0,830.0){\rule[-0.200pt]{0.400pt}{4.818pt}}
\put(392.0,150.0){\rule[-0.200pt]{0.400pt}{4.818pt}}
\put(392,100){\makebox(0,0){ 600}}
\put(392.0,830.0){\rule[-0.200pt]{0.400pt}{4.818pt}}
\put(525.0,150.0){\rule[-0.200pt]{0.400pt}{4.818pt}}
\put(525,100){\makebox(0,0){ 800}}
\put(525.0,830.0){\rule[-0.200pt]{0.400pt}{4.818pt}}
\put(658.0,150.0){\rule[-0.200pt]{0.400pt}{4.818pt}}
\put(658,100){\makebox(0,0){ 1000}}
\put(658.0,830.0){\rule[-0.200pt]{0.400pt}{4.818pt}}
\put(792.0,150.0){\rule[-0.200pt]{0.400pt}{4.818pt}}
\put(792,100){\makebox(0,0){ 1200}}
\put(792.0,830.0){\rule[-0.200pt]{0.400pt}{4.818pt}}
\put(925.0,150.0){\rule[-0.200pt]{0.400pt}{4.818pt}}
\put(925,100){\makebox(0,0){ 1400}}
\put(925.0,830.0){\rule[-0.200pt]{0.400pt}{4.818pt}}
\put(225.0,150.0){\rule[-0.200pt]{168.630pt}{0.400pt}}
\put(925.0,150.0){\rule[-0.200pt]{0.400pt}{168.630pt}}
\put(225.0,850.0){\rule[-0.200pt]{168.630pt}{0.400pt}}
\put(225.0,150.0){\rule[-0.200pt]{0.400pt}{168.630pt}}
\put(50,500){\makebox(0,0){$A^{t}_{FB}$}}
\put(575,25){\makebox(0,0){$M_{t\bar{t}}~\text{(GeV)}$}}
\put(258,733){\makebox(0,0)[l]{$g^q_A= - g^t_A$}}
\put(633,354){\makebox(0,0)[r]{$g^q_V= g^t_V$}}
\put(658.0,354.0){\rule[-0.200pt]{28.908pt}{0.400pt}}
\put(225,403){\usebox{\plotpoint}}
\put(225.0,403.0){\rule[-0.200pt]{24.090pt}{0.400pt}}
\put(325.0,403.0){\rule[-0.200pt]{0.400pt}{16.381pt}}
\put(325.0,471.0){\rule[-0.200pt]{24.090pt}{0.400pt}}
\put(425.0,471.0){\rule[-0.200pt]{0.400pt}{22.404pt}}
\put(425.0,564.0){\rule[-0.200pt]{24.090pt}{0.400pt}}
\put(525.0,564.0){\rule[-0.200pt]{0.400pt}{14.695pt}}
\put(525.0,625.0){\rule[-0.200pt]{24.090pt}{0.400pt}}
\put(625.0,567.0){\rule[-0.200pt]{0.400pt}{13.972pt}}
\put(625.0,567.0){\rule[-0.200pt]{24.090pt}{0.400pt}}
\put(725.0,405.0){\rule[-0.200pt]{0.400pt}{39.026pt}}
\put(725.0,405.0){\rule[-0.200pt]{24.090pt}{0.400pt}}
\put(825.0,243.0){\rule[-0.200pt]{0.400pt}{39.026pt}}
\put(825.0,243.0){\rule[-0.200pt]{24.090pt}{0.400pt}}
\sbox{\plotpoint}{\rule[-0.500pt]{1.000pt}{1.000pt}}%
\sbox{\plotpoint}{\rule[-0.200pt]{0.400pt}{0.400pt}}%
\put(633,304){\makebox(0,0)[r]{$g^q_V= - g^t_V$}}
\sbox{\plotpoint}{\rule[-0.500pt]{1.000pt}{1.000pt}}%
\multiput(658,304)(20.756,0.000){6}{\usebox{\plotpoint}}
\put(778,304){\usebox{\plotpoint}}
\put(225,404){\usebox{\plotpoint}}
\multiput(225,404)(20.756,0.000){5}{\usebox{\plotpoint}}
\multiput(325,404)(0.000,20.756){4}{\usebox{\plotpoint}}
\multiput(325,473)(20.756,0.000){4}{\usebox{\plotpoint}}
\multiput(425,473)(0.000,20.756){5}{\usebox{\plotpoint}}
\multiput(425,577)(20.756,0.000){5}{\usebox{\plotpoint}}
\multiput(525,577)(0.000,20.756){6}{\usebox{\plotpoint}}
\multiput(525,690)(20.756,0.000){5}{\usebox{\plotpoint}}
\multiput(625,690)(0.000,20.756){4}{\usebox{\plotpoint}}
\multiput(625,772)(20.756,0.000){4}{\usebox{\plotpoint}}
\put(725.00,775.73){\usebox{\plotpoint}}
\multiput(725,792)(20.756,0.000){5}{\usebox{\plotpoint}}
\multiput(825,792)(0.000,-20.756){2}{\usebox{\plotpoint}}
\multiput(825,754)(20.756,0.000){5}{\usebox{\plotpoint}}
\put(925,754){\usebox{\plotpoint}}
\sbox{\plotpoint}{\rule[-0.200pt]{0.400pt}{0.400pt}}%
\put(225.0,150.0){\rule[-0.200pt]{168.630pt}{0.400pt}}
\put(925.0,150.0){\rule[-0.200pt]{0.400pt}{168.630pt}}
\put(225.0,850.0){\rule[-0.200pt]{168.630pt}{0.400pt}}
\put(225.0,150.0){\rule[-0.200pt]{0.400pt}{168.630pt}}
\end{picture}
\caption{$A^{t}_{FB}$ vs. $M_{t\bar{t}}$ with $M_{t\bar{t}}$ integrated
over each 150 GeV of $M_{t\bar{t}}$
using the benchmark point, $M_{G}=1525~$GeV, $g^q_V=-0.577g_s$ 
and $g^q_A=-g^t_A=-1.155g_s$. 
The solid line corresponds to our model with $g^{q}_{V}=g^{t}_{V}$
and dashed line is a comparison plot with $g^{q}_{V}=-g^{t}_{V}$,
as explained in the text.}
\label{Fig:afbmtt}
\end{figure}

To test the competition effects, we propose to use the $A^t_{FB}$ dependence 
on $\sqrt{\hat{s}}$ $(=M_{t\bar{t}})$.  
To illustrate this feature of our model, we choose a benchmark point
within the 1~$\sigma$ region as $M_G = 1525$~ GeV
and $\theta=27^{\circ}$.
We plot (solid line) the $A^{t}_{FB}$ vs. $M_{t\bar{t}}$ with $M_{t\bar{t}}$ integrated
every each 150 GeV of $M_{t \bar{t}}$ in Fig. \ref{Fig:afbmtt}.
One can see the competition in $A^{t}_{FB}$ between two 
terms at different $M_{t\bar{t}}$ in Fig.\ref{Fig:afbmtt}
and when $M_{t\bar{t}}$ is close to on-shell $G^{1}$
region, the new physics contribution begins to dominate
over the interference term which bend over the curve and generate a negative
asymmetry. For comparison, we plot (in dashed line)
a case where  $g^{q}_{V}=-g^{t}_{V}$ so the new physics term also generates positive asymmetry. 

In other theoretical proposals  
and the QCD corrections \cite{Almeida:2008ug}, the $A^{t}_{FB}$ induced by other types of $s$-channel new physics without the above competition or $t$-channel new physics \cite{Jung:2009jz} grows with $M_{t\bar{t}}$ which is very different from our prediction. When the $A^{t}_{FB}$ vs $M_{t\bar{t}}$ 
distribution is available in the near future, it is then easy to confirm or rule out our model. This becomes crucial since the resonance is too heavy and cannot be directly produced at the Tevatron.

\section{Discussion}

The above minimal setup can be extended to an extra dimension scenario on an 
interval bounded by left and right branes. We allow all fermions 
to propagate in the bulk and interactions to be 
determined through the five dimensional wavefunction overlap. In the spirit of deconstruction \cite{ArkaniHamed:2001ca}, 
one can imagine that the left-handed light quarks and the right handed top quarks are localized on the left brane while the right-handed light quarks and the left handed top quarks are localized on right one. 
The fourth generation is not needed in this case because the gauge anomaly will be cancelled among each generation \cite{ArkaniHamed:2001is} \footnote[4]{If all fermions are completely localized at the boundary brane, the gauge anomaly is not cancelled among each generation and one need the fourth generation or a $SU(3)$ Chern-Simons term in the bulk. In this case, one can use the path ordering of the Wilson line to generate the quark masses.}. The axigluon is the 1st KK gluon ($g^1$) whose wave 
function contains a node, and changes sign from one side of the extra dimension to the other. 
As a result the $g^1$ coupling for the left localized fermions has a minus sign relative to the one for the right localized fermions. Such a localization pattern guarantees Eqs. (\ref{eq: condition}) which provide the essential requirements to fit the data and make the striking prediction. 
We expect that such a simple localization pattern can be embedded in different extra dimensional models \cite{Agashe:2003zs, Park:2009cs}. 

In this paper, we present a flavor-nonuniversal chiral color model to explain
the Tevatron anomaly on top quark forward-backward asymmetry. The model
predicts a heavy axigluon that couples to first two generation and
third generation quarks with opposite axial couplings but same vector couplings with relative small sizes. The data are consistent with our model within 1$\sigma$ for some parameters, and a large region of parameter space is found at $90\%$ C. L. One can use $A^{t}_{FB}$ vs $M_{t\bar{t}}$ distribution to further distinguish this model from other types of models that can explain $A^{t}_{FB}$ anomaly before one discovers the axigluon resonance. 

If the striking rise and fall behavior of $A^{t}_{FB}$ vs $M_{t\bar{t}}$ were discovered, 
the implication is not only that strong interactions will violate parity at much higher 
energies as originally motivated Ref. \cite{FG}, but Eqs. (\ref{eq: condition}) suggest
a possible specific pattern of parity violation.

\section{Notes Added}
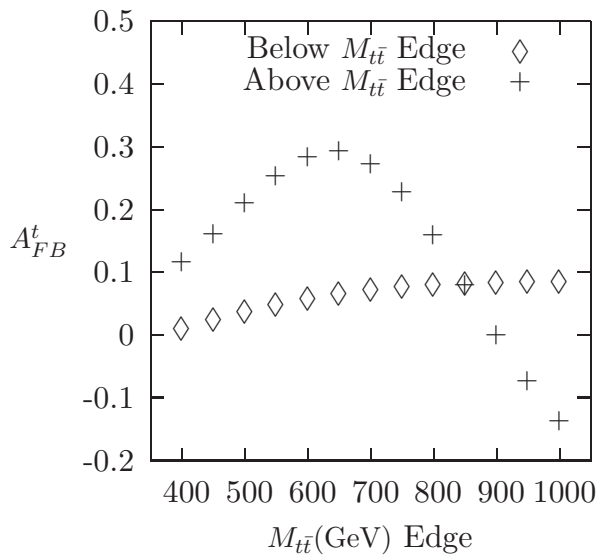
\begin{figure}[t]
\setlength{\unitlength}{0.240900pt}
\ifx\plotpoint\undefined\newsavebox{\plotpoint}\fi
\sbox{\plotpoint}{\rule[-0.200pt]{0.400pt}{0.400pt}}%
\begin{picture}(1500,900)(0,0)
\font\gnuplot=cmr10 at 12pt
\gnuplot
\sbox{\plotpoint}{\rule[-0.200pt]{0.400pt}{0.400pt}}%
\put(504.0,160.0){\rule[-0.200pt]{4.818pt}{0.400pt}}
\put(479,160){\makebox(0,0)[r]{-0.2}}
\put(1175.0,160.0){\rule[-0.200pt]{4.818pt}{0.400pt}}
\put(504.0,259.0){\rule[-0.200pt]{4.818pt}{0.400pt}}
\put(479,259){\makebox(0,0)[r]{-0.1}}
\put(1175.0,259.0){\rule[-0.200pt]{4.818pt}{0.400pt}}
\put(504.0,357.0){\rule[-0.200pt]{4.818pt}{0.400pt}}
\put(479,357){\makebox(0,0)[r]{ 0}}
\put(1175.0,357.0){\rule[-0.200pt]{4.818pt}{0.400pt}}
\put(504.0,456.0){\rule[-0.200pt]{4.818pt}{0.400pt}}
\put(479,456){\makebox(0,0)[r]{ 0.1}}
\put(1175.0,456.0){\rule[-0.200pt]{4.818pt}{0.400pt}}
\put(504.0,554.0){\rule[-0.200pt]{4.818pt}{0.400pt}}
\put(479,554){\makebox(0,0)[r]{ 0.2}}
\put(1175.0,554.0){\rule[-0.200pt]{4.818pt}{0.400pt}}
\put(504.0,653.0){\rule[-0.200pt]{4.818pt}{0.400pt}}
\put(479,653){\makebox(0,0)[r]{ 0.3}}
\put(1175.0,653.0){\rule[-0.200pt]{4.818pt}{0.400pt}}
\put(504.0,751.0){\rule[-0.200pt]{4.818pt}{0.400pt}}
\put(479,751){\makebox(0,0)[r]{ 0.4}}
\put(1175.0,751.0){\rule[-0.200pt]{4.818pt}{0.400pt}}
\put(504.0,850.0){\rule[-0.200pt]{4.818pt}{0.400pt}}
\put(479,850){\makebox(0,0)[r]{ 0.5}}
\put(1175.0,850.0){\rule[-0.200pt]{4.818pt}{0.400pt}}
\put(553.0,160.0){\rule[-0.200pt]{0.400pt}{4.818pt}}
\put(553,110){\makebox(0,0){ 400}}
\put(553.0,830.0){\rule[-0.200pt]{0.400pt}{4.818pt}}
\put(652.0,160.0){\rule[-0.200pt]{0.400pt}{4.818pt}}
\put(652,110){\makebox(0,0){ 500}}
\put(652.0,830.0){\rule[-0.200pt]{0.400pt}{4.818pt}}
\put(751.0,160.0){\rule[-0.200pt]{0.400pt}{4.818pt}}
\put(751,110){\makebox(0,0){ 600}}
\put(751.0,830.0){\rule[-0.200pt]{0.400pt}{4.818pt}}
\put(850.0,160.0){\rule[-0.200pt]{0.400pt}{4.818pt}}
\put(850,110){\makebox(0,0){ 700}}
\put(850.0,830.0){\rule[-0.200pt]{0.400pt}{4.818pt}}
\put(948.0,160.0){\rule[-0.200pt]{0.400pt}{4.818pt}}
\put(948,110){\makebox(0,0){ 800}}
\put(948.0,830.0){\rule[-0.200pt]{0.400pt}{4.818pt}}
\put(1047.0,160.0){\rule[-0.200pt]{0.400pt}{4.818pt}}
\put(1047,110){\makebox(0,0){ 900}}
\put(1047.0,830.0){\rule[-0.200pt]{0.400pt}{4.818pt}}
\put(1146.0,160.0){\rule[-0.200pt]{0.400pt}{4.818pt}}
\put(1146,110){\makebox(0,0){ 1000}}
\put(1146.0,830.0){\rule[-0.200pt]{0.400pt}{4.818pt}}
\put(504.0,160.0){\rule[-0.200pt]{0.400pt}{166.221pt}}
\put(504.0,160.0){\rule[-0.200pt]{166.462pt}{0.400pt}}
\put(1195.0,160.0){\rule[-0.200pt]{0.400pt}{166.221pt}}
\put(504.0,850.0){\rule[-0.200pt]{166.462pt}{0.400pt}}
\put(329,505){\makebox(0,0){$A^{t}_{FB}$}}
\put(849,35){\makebox(0,0){$M_{t\bar{t}} \text{(GeV)}$ Edge}}
\put(1000,805){\makebox(0,0)[r]{Below $M_{t\bar{t}}$ Edge}}
\put(553,370){\raisebox{-.8pt}{\makebox(0,0){$\Diamond$}}}
\put(603,384){\raisebox{-.8pt}{\makebox(0,0){$\Diamond$}}}
\put(652,397){\raisebox{-.8pt}{\makebox(0,0){$\Diamond$}}}
\put(701,408){\raisebox{-.8pt}{\makebox(0,0){$\Diamond$}}}
\put(751,418){\raisebox{-.8pt}{\makebox(0,0){$\Diamond$}}}
\put(800,426){\raisebox{-.8pt}{\makebox(0,0){$\Diamond$}}}
\put(850,432){\raisebox{-.8pt}{\makebox(0,0){$\Diamond$}}}
\put(899,436){\raisebox{-.8pt}{\makebox(0,0){$\Diamond$}}}
\put(948,439){\raisebox{-.8pt}{\makebox(0,0){$\Diamond$}}}
\put(998,441){\raisebox{-.8pt}{\makebox(0,0){$\Diamond$}}}
\put(1047,443){\raisebox{-.8pt}{\makebox(0,0){$\Diamond$}}}
\put(1096,444){\raisebox{-.8pt}{\makebox(0,0){$\Diamond$}}}
\put(1146,445){\raisebox{-.8pt}{\makebox(0,0){$\Diamond$}}}
\put(1085,805){\raisebox{-.8pt}{\makebox(0,0){$\Diamond$}}}
\put(1000,755){\makebox(0,0)[r]{Above $M_{t\bar{t}}$ Edge}}
\put(553,472){\makebox(0,0){$+$}}
\put(603,517){\makebox(0,0){$+$}}
\put(652,565){\makebox(0,0){$+$}}
\put(701,608){\makebox(0,0){$+$}}
\put(751,637){\makebox(0,0){$+$}}
\put(800,646){\makebox(0,0){$+$}}
\put(850,627){\makebox(0,0){$+$}}
\put(899,582){\makebox(0,0){$+$}}
\put(948,514){\makebox(0,0){$+$}}
\put(998,436){\makebox(0,0){$+$}}
\put(1047,357){\makebox(0,0){$+$}}
\put(1096,285){\makebox(0,0){$+$}}
\put(1146,223){\makebox(0,0){$+$}}
\put(1085,755){\makebox(0,0){$+$}}
\put(504.0,160.0){\rule[-0.200pt]{0.400pt}{166.221pt}}
\put(504.0,160.0){\rule[-0.200pt]{166.462pt}{0.400pt}}
\put(1195.0,160.0){\rule[-0.200pt]{0.400pt}{166.221pt}}
\put(504.0,850.0){\rule[-0.200pt]{166.462pt}{0.400pt}}
\end{picture}
\caption{$A^{t}_{FB}$ vs. above/below $M_{t\bar{t}}$ edge using the benchmark point, $M_{G}=1525~$GeV, 
 $g^q_A=-g^t_A=-1.155g_s$ and $g^{q}_{V}=g^{t}_{V}=-0.577g_s$ 
}
\label{Fig:afbmtt2}
\end{figure}

After submission of our paper, we notice that the CDF collaboration measures 
the correlation between $A^{t}_{FB}$ and $M_{t\bar{t}}$
by selecting events below or above $M_{t\bar{t}}$ edge point. 
To make a direct comparison with the CDF public result \cite{cdfmttafb}, 
we also plot $A^{t}_{FB}$ vs $M_{t\bar{t}}$ edge
distributions in Fig.\ref{Fig:afbmtt2}. This plot can be easily 
compared to the corresponding 
figure in \cite{cdfmttafb}.

\section{Acknowledgement}
This work was supported by the World Premier International Research Center Initiative 
(WPI initiative) MEXT, Japan. The work of P.H.F. was also supported by U.S. Department
of Energy Grant No. DE-FG02-05ER41418. The work of J.S. was also supported by
the Grant-in-Aid for scientific research (Young Scientists (B)
21740169) from JSPS.

\end{document}